# A "Muon Mass Tree" with α-quantized Lepton, Quark and Hadron Masses

Malcolm H. Mac Gregor
130 Handley Street, Santa Cruz. CA 95060
July 20, 2006
e-mail: mhmacgreg@aol.com;  web-site: 70mev.org

**Abstract**

One of the most important unfulfilled tasks in elementary particle physics is to develop a comprehensive mass formalism that encompasses leptons, quarks and hadrons. The Standard Model (SM) treatment of particle masses, which places leptons and hadrons in separate unrelated categories, has been admittedly unsuccessful. However, by combining the mass data for leptons and hadrons, and by including particle lifetimes in the analysis, we arrive at an overall mass formalism based on experiment that accurately applies to leptons, quarks, hadrons and gauge bosons. There are five salient features in this formalism:

  *(1)* all types of particles, including quark substates, share the same basic mass units,
     so their various interaction modes are attributable to their electric charge states;
  *(2)* the long-lived ($\tau > 10^{-21}$ sec) particles have α-scaled lifetimes and reciprocal
     $\alpha^{-1}$-scaled masses, where $\alpha = e^2/\hbar c \cong 1/137$ is the fine structure constant;
  *(3)* particle masses are generated by the action of the coupling constant α on the electron
     ground state, and also on the proton $u$ and $d$ quarks in Tevatron high-energy collisions;
  *(4)* α-generated mass quanta are the *boson* mass $m_b = m_e/\alpha \cong 70$ MeV and the *fermion*
     masses $m_f = (3/2)(m_e/\alpha) \cong 105$ MeV and $m_f^\alpha = (3/2)(m_e/\alpha^2) \cong 14{,}394$ MeV.
  *(5)* an unbreakable 2-3% hadronic binding energy (HBE) applies to quark-antiquark pairs
     below 6 GeV, and is negligible for higher energy pairs and unpaired states, so the SM
     $u, d, s, c, b, t$ quarks combine together with additive *constituent-quark* masses.

The mass quanta $m_b$ and $m_f$ reproduce the elementary particle states below 12 GeV, and the mass quantum $m_f^\alpha \equiv m_f/\alpha$ reproduces the particle states W, Z and $t$ above 12 GeV. Evidence for the universality of these mass quanta is provided by the fact that the sum of the *gauge boson* $W^\pm$ and $Z^o$ masses is equal to the *top quark t* mass (to 0.5% accuracy). Patterns for the way these α-quantized mass quanta combine together in the basic threshold states are shown in the form of a "muon mass tree" composed of additive $m_f$ and $m_f^\alpha$ mass units and a "pion mass tree" composed of $m_b$ mass units. The muon mass tree reproduces the μ and τ lepton masses, the $u, d, s, c, b, t$ constituent-quark masses, the proton mass, the φ, J/ψ$_{1S}$ and ϒ$_{1S}$ vector meson masses, the B$_c$ meson mass, and the $\overline{WZ}$ average mass to an overall mass accuracy of 0.43%, using no adjustable parameters except a 2.6% HBE for the φ and J/ψ$_{1S}$ states. The pion mass tree reproduces the isotopic-spin-averaged π, η, η' and K pseudoscalar meson masses to an accuracy of 0.38% when a 2.6% HBE is applied to the paired π, η and η' states.

———

The mu meson, or *muon*, has had a long and unfortunate history. At first it was thought to be the particle predicted by Yukawa that mediates the strong (hadronic) and short-ranged nuclear force, since it has the predicted mass. However, its interactions turned out to be weak (leptonic) and long-ranged. This dilemma was resolved by the subsequent discovery of the pi meson, or *pion*, which has roughly the same mass as the muon, and which does interact strongly (although we no longer regard the pion as the generator of the hadronic force). The pion was hailed as the Yukawa particle, and it heralded the appearance of a whole family of hadrons. The muon was left as a particle without a purpose—a leptonic outcast. This situation was famously summarized by I. I. Rabi in his rhetorical question about the muon:

"Who ordered that?"

A half century later, Frank Wilczek updated this situation as follows [1]:

"Quark and lepton masses ... have eluded calculation despite decades of intense effort. ... When the muon was discovered, I. I. Rabi asked, 'Who ordered that?' His question continues to resonate, undamped, with ever-increasing amplitude."

This recent assessment by Wilczek contains two interesting elements: *(1)* the existing theories of elementary particle masses do not encompass either the quarks or the leptons, and thus seem manifestly incomplete—a crucial element is missing; *(2)* the observed elementary particle states are almost all unstable particles, and yet the longest-lived (except for the neutron) and lowest-mass unstable state—the muon—which should be one of the most important particles, doesn't seem to fit in, and logic demands "with ever-increasing amplitude" that it *ought* to fit in. Hence we have on the one hand a particle mass theory with a missing element, and on the other hand an important particle with no theory to make use of it. These two related difficulties suggest the following joint solution:

*The muon is the missing element of the mass theories!*

A specific difficulty with the existing mass theories has recently been emphasized by Harald Fritsch [2]:

"We know the Standard Model does not address a number of decisive problems, most urgently the question for the origin of particle masses. The most telling of these, as far as the structure of matter is concerned, are the electron mass of 0.511 MeV and the proton mass of 938 MeV."

The challenge here, as first offered by Rabi, and then forcefully restated many years later by Wilczek, and augmented by Fritsch, is to provide a rationalé for the existence of the muon. The purpose of the present paper is to demonstrate, directly from the experimental data, that by adding one new facet to the present elementary particle phenomenology, we can reveal the key role played by the muon mass in reproducing the mass values of the *u, d, s, c, b, t* Standard Model quarks (which appear here as *constituent* quarks), and also the mass values of the basic fermionic particle states—the heavy leptons and the proton. This new phenomenological element is the well-documented [3] quantization of the long-lived elementary particle lifetimes in powers of the fine structure constant $\alpha \cong 1/137$, as we now demonstrate.

There are 36 unstable elementary particles that have well-determined *mean lives* or *lifetimes* $\tau$ which are longer than $10^{-21}$ second [4]. These lifetimes, which span 23 orders of magnitude, are displayed in Fig. 1(a), where they are identified as to their boson or fermion spin nature and their dominant Standard Model quark content. As can be seen, they fall into well-defined lifetime groups which are each dominated by a single quark state. The precise nature of the lifetime group spacing is revealed in Fig. 1(b), where a factor-of-two "hyperfine" (HF) structure has been removed, and where the lifetimes are expressed as ratios to the $\pi^{\pm}$ reference lifetime and plotted as logarithms to the base $\alpha = e^2/\hbar c$, as follows:

$$\tau_i = \tau_{\pi^{\pm}} \alpha^{x_i},$$

where the fine structure constant $\alpha \cong 1/137$ serves as the scaling factor. As can be seen in Fig. 1(b), the unpaired-quark lifetime logarithms $x_i$ have almost integer values except for the *c*-quark lifetimes (green), which are characteristically a factor of 3 shorter than the *b*-quark lifetimes (red). (Also, *c*-quark masses are a factor of 3 smaller than *b*-quark masses.) This lifetime systematics is analyzed in considerable detail in a recent invited review paper [5], and also in a forthcoming book [6].

The crucial aspect of the lifetime plots of Fig. 1 for our present purpose is that they exhibit a clear-cut dependence on the constant $\alpha$. These 36 long-lived particle states correspond to the *thresholds* where the various quark combinations first occur, and they are the states that are the most important with respect to the mass systematics. Since particle *lifetimes* or *mean lives* are reciprocally related by the uncertainty principle to particle *mass widths* [4], which in turn are logically related to particle *masses*, an $\alpha$-dependence in the lifetimes should be accompanied by a corresponding $\alpha^{-1}$-dependence in the masses. This is a straightforward hypothesis to investigate experimentally. Threshold-state lifetimes are observed to decrease by successive factors of $\alpha \cong 1/137$, and threshold-state masses should thereby increase by mass units that contain factors of $\alpha^{-1} \cong 137$. To see if this is actually the case, we show in Fig. 2 a plot of the masses of these 36 long-lived unstable particles, together with the masses of the stable electron and proton. If we do not include the electron, there are no factor-of-137 mass ratios in Fig. 2. However, with the electron added in as a member of this group, two prominent $\alpha^{-1}$-quantized mass units emerge—the bosonic mass quantum $m_b \cong 70$ MeV, and the fermionic mass quantum $m_f \cong 105$ MeV, as diagrammed in Fig. 2. If we treat $m_f$ as an excitation mass that adds to the electron mass, then we obtain the mass equation $m_\mu = m_e + m_f$, which reproduces the muon mass to 0.1% accuracy. As we will now demonstrate, the muon mass $m_e + m_f$ acts as a *platform* state for higher-mass excitations: multiples of the basic fermion mass $m_f$ are added to the platform mass to create the branches of a "muon mass tree" that accurately reproduces the SM quarks and the fundamental leptons and hadrons, with the fermion "$\alpha$ mass" $m_f^\alpha = m_f/\alpha$ extending these results up to the mass region above 12 GeV.

Fig. 3 displays the *muon mass tree*, which is constructed out of the first-order and second-order $\alpha$-quantized masses $m_f \cong 105$ MeV and $m_f^\alpha \cong 14{,}394$ MeV, plus one or two ground-state electron masses. Table 1 presents these same results in a tabular form. The first excited state is the muon itself, which serves as a platform for creating higher-mass levels. A series of additional excited states are formed by four successive doublings of the excitation quantum $2\,m_f \cong 210$ MeV, which add to the muon platform mass to generate *(1)* the isotopic-spin-averaged $q \equiv (u,d)$ nucleon quark constituent mass, *(2)* the $s$-quark constituent mass, *(3)* the proton mass (0.8% accuracy), and *(4)* the tau lepton mass (0.5% accuracy). The $s$ quark mass is then tripled twice to produce the $c$ and $b$ quark constituent masses. All of these excitations are carried out in parallel particle and antiparticle excitation channels. The $c$ and $b$ quarks combine additively to reproduce the $B_c \equiv \bar{b}c$ meson mass (0.3% accuracy) and the $\Upsilon_{1S} \equiv b\bar{b}$ vector meson mass (0.1% accuracy). The $q$ and $s$ quark mass excitations are accompanied by charge fractionation (CF) into 1/3 and 2/3 fragments, with this CF process carrying over to the $c$ and $b$ quarks. The $s$ to $c$ to $b$ quark mass triplings are also accompanied by cross-channel charge exchange (CX) processes that result in the −1/3 to +2/3 to −1/3 charge alternations in the quark states. The creation of a proton-antiproton pair is accompanied by a double CX cross-channel charge transfer that leaves the positively-charged proton trapped in the negatively-charged particle channel. These low-mass (< 12 GeV) muon mass tree excitation sequences, which are based on an initial $m_f$ α-leap from the electron ground state plus additional $m_f$ excitation quanta, are sufficient to satisfy the main thrust of the concerns voiced by Rabi, Wilczek and Fritsch [1,2]. The "muon-like" fermion mass quantum $m_f$ = 105.038 MeV is the fundamental mass unit for producing all of these states. Since the Standard Model hadronic mass formalism does not include this mass unit, it cannot reproduce the results displayed in the muon mass tree. It should be noted that no freely-adjustable parameters are used in these accurate additive mass fits. For higher mass accuracy, the paired-quark states below 6 GeV (the ϕ and J/ψ mesons in Fig. 3) require a small (2-3%) hadronic binding energy correction (see Fig. 7 and Table 1), but this has not been applied to the mass calculations of Fig. 3.

The possibility of extending these low-mass muon tree results upward to also include the high-mass states well above 12 GeV was suggested by an experimental fact which has largely escaped the attention of the Standard Model community: the (W + Z) to $t$ mass ratio. There are presently only three accurately-measured particle states in the high-mass region: the W and Z gauge bosons, and the top quark $t$. The W and Z masses are [4]:

$$m_{W^\pm} = 80.425 \pm 0.038 \text{ GeV}, \quad m_{Z^o} = 91.1876 \pm 0.0021 \text{ GeV}. \qquad (1)$$

The recent Tevatron D0-CDF consensus mass value for the $t$ mass is [7]

$$m_t = 172.5 \pm 2.3 \text{ GeV}. \qquad (2)$$

As can be seen, all three of these masses are known very accurately. According to the current Standard Model paradigm, there is no reason to expect the W and Z gauge boson masses to bear any relationship to the $t$ quark mass. But experimentally we have the mass relationship

$$m_W + m_Z = 171.61 \pm 0.04 \text{ GeV}, \quad m_t = 172.5 \pm 2.3 \text{ GeV}. \qquad (3)$$

This is mass agreement to an accuracy of 0.5%, which is within the experimental errors on the mass values. If we decide that the mass accuracy of this result, and also its comprehensiveness (since these are the only known particles in this mass region), cannot be accidental, then it leads to an important phenomenological conclusion: the W and Z gauge boson masses are in some sense related to the $t$ quark mass, which in turn is logically related to the $b$ quark mass— it's SM partner. Hence these very massive states should tie in with the mass systematics of the lower-mass states.

The clue to the mass value of the $t$ quark comes from its production process in the Tevatron. The $t$ quark is produced in $t\bar{t}$ pairs by proton-antiproton collisions at the highest available Tevatron energies. At these energies, the collisions actually occur between the individual $q \equiv (u, d)$ and $\bar{q} \equiv (\bar{u}, \bar{d})$ quarks in these particles. Thus the $q$ quarks $u$ and $d$ logically form the "ground states" for the $t$ quark excitations. A $t\bar{t}$ pair production event requires a head-on quark-antiquark collision. This event is identified and measured in the exclusive four-prong decay channel $t \to W^+ + b$, $\bar{t} \to W^- + \bar{b}$, where the $t$ and $\bar{t}$ decay streams each feature two decay centers—a prompt W decay and a delayed $b$ decay. These events occur once in every $10^{10}$ proton-antiproton collisions. The mass leap upward from the 315 MeV $q$ quarks to the 172,500 MeV $t$ quark is more than two orders of magnitude, which suggests a large mass increase of the $q$ quarks in a single step. This indicates an $\alpha$-enhancement ($\alpha$-leap) of the $q$ quark mass, which itself was created by an $\alpha$-enhancement of the electron mass (Figs. 2 and 3). The mass equation for the initial single-$\alpha$-leap creation of a $q$ quark is (Table 1)

$$m_q \equiv (u,d) = m_e(1+9/2\alpha) = 315.625 \text{ MeV}. \qquad (4)$$

The equation for a second $\alpha$-enhancement up to a much-higher-mass "$\alpha$-quark" pair $q^\alpha$, which is two $\alpha$-leaps above the electron ground state, is

$$m_{q^\alpha} \equiv (m_{u^\alpha}, m_{d^\alpha}) = m_e(1+9/2\alpha^2) = 43{,}182.5 \text{ MeV}. \qquad (5)$$

The $q^\alpha$ $\alpha$-quarks serve as new high-mass basis states. The W and Z bosons are reproduced as $q^\alpha \bar{q}^\alpha$ $\alpha$-quark pairs, whose calculated isotopic-spin-averaged mass value is

$$(m_{\overline{WZ}})_{\text{calc}} \equiv (\overline{WZ})_{\text{calc}} = m_{q^\alpha \bar{q}^\alpha} = m_e(2+9/\alpha^2) = 86.365 \text{ GeV}, \qquad (6)$$

which is within 0.65% of the experimental mass

$$(\overline{WZ})_{\text{exp}} = {} = 85.806 \text{ GeV}. \qquad (7)$$

Furthermore, the $q^\alpha \bar{q}^\alpha$ quark pairs have the charge freedom to reproduce the $W^\pm$ and $Z^o$ isotopic-spin states. The experimental mass relationship $m_t = 2m_{\overline{WZ}}$ (Eq. 3) leads in turn to the quark assignment $t \Leftrightarrow 4q^\alpha$. Hence the calculated $t$ quark mass is [8]

$$(m_t)_{\text{calc}} = 4m_{q^\alpha} = m_e(1+18/\alpha^2) = 172.73 \text{ GeV}, \qquad (8)$$

which accurately matches the experimental value,

$$(m_t)_{\exp} = 172.5 \pm 2.3 \text{ GeV}. \qquad (9)$$

Hence the same basic fermion mass α-generation process applies to both the lower-mass particle states below 12 GeV, which correspond to a single α-leap from an electron ground state, and also the higher-mass states above 12 GeV, which correspond to a second α-leap from an $m_f$ or $q = 3m_f$ ground state.

The mass systematics displayed in Fig. 3 and tabulated in Table 1 is sufficient to define the quark masses for the six Standard Model quarks, and to reproduce the lepton and proton masses, the $s\bar{s}$, $c\bar{c}$ and $b\bar{b}$ vector meson threshold states, the $B_c$ meson mass, and the $\overline{WZ}$ average mass. The particle masses are accurately reproduced by simply adding up the quark masses. Thus it is apparent that this is a *constituent-quark* formalism, wherein the intrinsic quark masses determine the masses of the observed states. It then follows that the 315 MeV $q \equiv (u,d)$ quarks which reproduce the proton cannot be used in this same manner to reproduce the 140 MeV pions and other pseudoscalar (PS) mesons, as is done in the Standard Model (where the $u$ and $d$ quarks are assigned "current-quark" masses of just a few MeV). These low-mass spin 0 PS mesons require the "pion mass tree", which we now discuss.

Fig. 4 displays the spin 0 *pion mass tree*, which is constructed out of the first-order α-quantized boson masses $m_b \cong 70$ MeV that are diagrammed in Fig. 2. These results are also included in Table 1. The pion mass tree reproduces the isotopic-spin-averaged mass values of the π, η, η', K meson nonet. The "quark" states $q_\pi$, $q_\eta$, $q_K$ shown in Fig. 4 represent combinations of 1, 4, 7 $m_b$ mass quanta, respectively. They are formed by successive doublings of the excitation quantum 3 $m_b \cong 210$ MeV, thus echoing the 2 $m_f \cong 210$ MeV excitation doubling displayed in the muon mass tree of Fig. 3. These are "generic" quarks, whose main function here is to reproduce mass values. The "unpaired" kaon isotopic spin states have a calculated average mass value that is equal to $m_{q_K}$ (1% accuracy). The "paired-quark" states π, η and η' have calculated masses that are all too large by about 2.6% (Table 1), and thus require HBE corrections (Fig. 7). The pion quarks $q_\pi$, $q_\eta$, and $q_K$ are *constituent* quarks, which are required to reproduce the basic low-mass threshold states (the PS nonet) that cannot be reproduced by the muon constituent quarks *u*, *d*, *s*, *c* and *b*. Pionic mass units also appear in "hybrid" excitations that are reproduced as combinations of muon and pion basis states, as discussed in Refs. [5] and [6].

Figs. 5 and 6 are slightly different displays of the muon and pion mass trees of Figs. 3 and 4, and are designed to bring out the universality of the $m_f$, $m_f^\alpha$ and $m_b$ mass quantizations. The Fig. 5 muon mass tree contains the quark states

$$(q, s, c, b) = (3, 5, 15, 45)m_f, \quad (q^\alpha, t) = (3, 12)m_f^\alpha,$$

and the particle states

$$(\mu, p, \tau) = (1, 9, 17)m_f, \quad (\phi, J/\psi, B_c, \Upsilon) = (5, 15, 30, 45)m_f \bar{m}_f, \quad \overline{WZ} = 3m_f^\alpha \bar{m}_f^\alpha.$$

The Fig. 6 pion mass tree contains the generic quark states

$$(q_\pi, q_\eta, q_K) = (1, 4, 7)m_b,$$

and the particle states

$$K = 7m_b, \quad (\pi, \eta, \eta') = (1, 4, 7)m_b \bar{m}_b.$$

(The various K meson configurations—$K^\pm, K_L^o, K_S^o$—may correspond to different combinations of particle and antiparticle subquanta $m_b$ and $\bar{m}_b$.)

If we determine the masses of these states by simply adding up their basis-state masses, plus one or two electron masses, as shown in Table 1, then the overall isotopic-spin-averaged mass accuracy for the nine particle states of the muon mass tree whose masses have been directly measured—μ, p, τ, ϕ, J/ψ, $B_c$, Υ, $\overline{WZ}$, *t*—is 0.83%, with no hadronic binding energy corrections applied. In detail, the three low-mass unpaired-quark particle states μ, p and τ, which require no HBE corrections, are reproduced to an average mass accuracy of 0.48%, and the four high-mass particle and quark states $B_c$, Υ, $\overline{WZ}$ and *t*, where the HBE seems to essentially vanish (Fig. 7), are reproduced to an average mass accuracy of 0.28%. The ϕ and J/ψ vector mesons have calculated HBE's of 3.1% and 1.8%, respectively (Table 1). The mass value of the one unpaired-quark state in the pion mass tree—the kaon K—is reaproduced to 1% accuracy, and the mass values of the three paired-quark states each reflect an HBE of about 2.6% (Fig. 7). The $m_b$, $m_f$ and $m_f^\alpha$ basis-state mass values are all determined from their α-quantizations (Table 1). Thus the only free parameter in these mass tree calcula-tions is the number of $m_b$, $m_f$ or $m_f^\alpha$ mass quanta to use in each case, and these numbers form clear-cut patterns that do not represent random choices. Hence we have a constituent-quark model that is essentially parameter-free.

The calculated hadronic binding energies of the π, η, η', ϕ, J/ψ, $B_c$ and ϒ paired-quark particles that are displayed in Figs. 3 - 6 and tabulated in Table 1 are plotted together in Fig. 7. Also included is the experimental $\overline{p}n$ binding energy of 4.4% [9], which is the largest binding energy ever measured. As can be seen in Fig. 7, the calculated HBE values are in the 2.5% - 3% range up to mass values of 1 Gev, and then decrease at the higher energies, and essentially disappear by an energy of 9 GeV, as discussed in the figure caption. This decrease in the HBE can be attributed to the decrease in the Compton radii of these particles with increasing mass, which brings the separated fractional charges closer together and produces the "asymptotic freedom" of the hadronic gluon forces. The calculated mass values of the very-high-mass $\overline{WZ}$ isospin doublet and the top quark $t$ (Table 1 and Fig. 3) are accurately obtained under the assumption of zero binding energy. The mass accuracies quoted in the muon and pion mass trees (and also listed in Table 1) were obtained by ignoring HBE corrections. If we for simplicity impose a uniform HBE of 2.6% on the paired-quark states π, η, η', ϕ and J/ψ, then all of the particle states shown in Figs. 3 and 4 and listed in Table 1 are reproduced at the 1% accuracy level.

The threshold-state particles considered here represent the most clear-cut examples of constituent-quark combinations: they are all pure muon or pure pion excitations. However, many hybrid muon-pion combinations exist, some of which can be accounted-for by their production channels. The quark and particle mass values obtained here, which are in the 1% accuracy range, are precise enough that deviations from calculated values can be quantitatively investigated, as we now briefly discuss. The proton mass is accurately given as the sum of three $q \equiv (u,d)$ quarks. However, the calculated $\Lambda = qqs$ and $\Xi = qss$ hyperon masses are each about 3.6% higher than expected, which suggests that hadronic binding energies may play a role in these excitations. The logical quark assignment for the lowest vector meson state is $q\bar{q}$, but the ρ and ω mesons appear at energies which are roughly 140 MeV higher, thus suggesting a $q\bar{q}\pi$ excitation. The quark assignment for the Ω hyperon is *sss*, but its observed mass (relative to the Λ and Ξ masses) is 140 MeV higher than calculated, which also suggests a hybrid excitation. The $\Sigma = qqs$ hyperon masses appear 70 MeV above the $\Lambda = qqs$ hyperon mass, and the lowest-mass spin 1 $K^* = q\bar{s}$ meson appears about 70 MeV above its expected value. These may also be hybrids. The short-lived ($\tau < 10^{-21}$ sec) particle excited states, which do not represent threshold "ground states" where the various quark combinations first appear, may in some cases represent "clustered" excitation units. These ramifications are discussed in more detail in Refs. [5] and [6].

As we have shown, the pion mass tree is based on the 70 MeV boson mass quantum $m_b$, and the muon mass tree is based on the 105 MeV fermion mass quantum $m_f$, which are displayed and defined in Fig. 2. Invoking the special-relativistic results [10] that *(1)* a fully relativistic spinning sphere (RSS) of matter is half again as massive as it was at rest, and *(2)* an RSS with a radius equal to its Compton radius has a calculated spin value of $1/2\ \hbar$, we can identify the α-quantized masses $m_b$ and $m_f$ as spin 0 and spin 1/2 configurations of the same basic 70 MeV mass quantum.

The W and Z gauge bosons and top quark $t$, which occupy the upper branches of the muon mass tree, have three features that are phenomenologically significant:

(1) They extend the domain of the muon mass quantum $m_f = 105$ MeV upward in energy, but not in the form of increased multiples of this mass unit. Instead, they occur as multiples of the "α-enhanced" mass quantum $m_f^\alpha \cong 137 \times 105$ MeV, which is a much larger but still "muon-based" excitation unit.

(2) They extend the elementary particle mass range upward so that it now encompasses two powers of $\alpha^{-1} \cong 137$ with respect to the electron ground state, and thus to an extent mirrors the reciprocal lifetime scaling, which extends over many powers of α. Additional *upward* mass scaling may be hard to obtain due to physical and fiscal limitations, but *downward* mass scaling into the realm of the neutrino masses is a possibility.

(3) The W, Z and $t$ particle and quark states represent "new physics", which arises due to their unanticipated and highly accurate mass equation $m_W + m_Z = m_t$. It has always been the hope (and the rationalé) of particle physicists that in building larger and more expensive accelerators, they will not only verify the already existing theories, but will also obtain some *new physics* results in the form of unexpected "surprises", which will then serve to guide them to more accurate formulations of their theories. The W, Z and $t$ mass relationship establishes a mass link between the W, Z *gauge boson* doublet and the *t quark*, which were thought to be unrelated entities. It also calls into question the nature of the $W^\pm, Z^o$ doublet itself. Should it be regarded as a $W^\pm, W^o$ isotopic spin doublet with a meaningful average mass value, as the present results suggest?

A final example that illustrates the usefulness of constituent quarks in representing elementary particle masses is the $B_c$ meson. This is the threshold state for the combination of a single *c*-quark with a single *b*-antiquark (or vice versa), and is at a high enough energy that binding energy effects are very small (Fig. 7). The experimental value of the $B_c$ mass, which was listed in Ref. [4] as 6400 MeV, has recently been lowered to the value [11]

$$(m_{B_c})_{exp} = 6287 \pm 4.8 \pm 1.1 \text{ MeV},$$

based on the measurement of the exclusive decay mode $B_c^\pm \to J/\psi^o \pi^\pm$. Shortly before this experimental result was announced, a large-scale lattice gauge theoretical collaboration between several groups published the following theoretical value for the $B_c$ mass [12]:

$$(m_{B_c})_{theor} = 6304 \pm 4 \pm 11 \text{ MeV}.$$

The agreement between these two results was properly hailed in the literature [13,14] as a triumph for lattice gauge calculations (and of course for the very accurate experimental measurement). The $B_c$ mass in the present constituent-quark formalism has the value (Table 1)

$$(m_{B_c})_{theor} = 2m_e + 60m_f = 6303.3 \text{ MeV}.$$

This constituent-quark value is in accurate agreement with the lattice-gauge value, which required cooperative world efforts on the largest obtainable computers. It should be noted that this constituent-quark value for the lowered $B_c$ mass was first published [15] well before the new experimental [11] and theoretical [12] $B_c$ mass values appeared. The lattice gauge calculations, of course, tie together many facets of particle physics, which are used to obtain values for the adjustable parameters of the formalism. But it is handy to also have a simple parameter-free calculational method available that yields accurate mass values and can serve as a guide for further refinements.

An independent experimental determination of constituent-quark masses is provided by the hyperon magnetic moments, which, if interpreted as Dirac point particles, yield the following quark mass values [16]: *u* = 338 MeV; *d* = 322 MeV; *s* = 510 MeV. The α-quantized

Table 1 values for these same quark states are: $\overline{u,d}$ = 315.6 MeV; $s$ = 525.7 MeV. These are very different ways of arriving at mass values for these quarks, so this close agreement should be taken seriously when judging the present results.

**References.**

Table 1. α-quantized mass values for the 13 quark and particle states displayed in the muon and pion mass trees of Figs. 3 and 4. Mass values are additive. With no binding energies (BE) applied, the average mass accuracy for these 13 particle states is 1.24%. With 2.6% BE's applied to 5 low-mass paired-quark states (see Fig. 7), the average 13-state mass accuracy is 0.41%.

*Input data:*

$m_e$ = 0.51099892 MeV

α=1/137.0359991

*First-order and second-order α-generated mass quanta:*

$m_b = m_e/\alpha$ = 70.025 MeV   (J = 0 boson)

$m_f = (3/2)(m_e/\alpha)$ = 105.038 MeV   (J = 1/2 femion)

$m_f^\alpha = (3/2)(m_e/\alpha^2)$ = 14,394.0 MeV   (J = 1/2 femion)

| Quark states | Mass units | Observed particle | Calculated mass | Exper. mass | Error (no BE) | Error (w BE) |
|---|---|---|---|---|---|---|
| *First-order $m_f$ excitation doubling* | | | | | | |
| μ lepton | $m_e + m_f$ | muon | 105.549 | 105.658 | -0.10% | |
| $q \equiv (u,d)$ quarks* | $\mu + 2m_f$ | | 315.625* | | | |
| s quark | $\mu + 4m_f$ | | 525.700 | | | |
| p = qqq* | $\mu + 8m_f$ | proton | 945.85* | 938.27 | +0.81% | |
| τ lepton | $\mu + 16m_f$ | tauon | 1786.16 | 1776.99 | +0.52% | |
| *First-order quark-mass tripling* | | | | | | |
| s quark | $m_e + 5m_f$ | | 525.700 | | | |
| c quark | $m_e + 15m_f$ | | 1576.079 | | | |
| b quark | $m_e + 45m_f$ | | 4727.215 | | | |
| $s\bar{s}$ | $2m_e + 10m_f$ | φ | 1051.40 | 1019.46 | +3.13% | +0.45% |
| $c\bar{c}$ | $2m_e + 30m_f$ | $J/\psi_{1S}$ | 3152.16 | 3096.92 | +1.78% | -0.87% |
| $c\bar{b}$ ** | $2m_e + 60m_f$ | $B_c$ ** | 6303.29** | 6287 | +0.26% | |
| $b\bar{b}$ | $2m_e + 90m_f$ | $\Upsilon_{1S}$ | 9454.43 | 9460.30 | –0.062% | |
| *First-order $m_f$ and second-order $m_f^\alpha$ mass triples* | | | | | | |
| q quarks* | $m_e + 3m_f$ | | 315.625* | | | |
| $q^\alpha \equiv (u^\alpha, d^\alpha)$ * | $m_e + 3m_f^\alpha$ | | 43,182.4* | | | |
| *Second-order "α-quark" particle states* | | | | | | |
| $q^\alpha \bar{q}^\alpha$ * | $2m_e + 6m_f^\alpha$ | $\overline{WZ}$ | 86,364.8* | 85,806.3* | +0.65% | |
| $4q^\alpha$ | $m_e + 12m_f^\alpha$ | t quark | 172,728.1 | 172,500 | +0.13% | |
| *First-order $m_b$ pseudoscalar mesons* | | | | | | |
| $q_\pi \bar{q}_\pi$ * | $2m_e + 2m_b$ | π | 141.07* | 137.27* | +2.77% | +0.10% |
| $q_\eta \bar{q}_\eta$ | $2m_e + 8m_b$ | η | 561.23 | 547.75 | +2.46% | -0.20% |
| $q_K \bar{q}_K$ | $2m_e + 14m_b$ | η' | 981.38 | 957.78 | +2.46% | -0.20% |
| $q_K$ * | $m_e + 7m_b$ | K | 490.69* | 495.66* | –1.00% | |

*Isotopic-spin-averaged masses   **Intermediate state in mass-tripling sequence

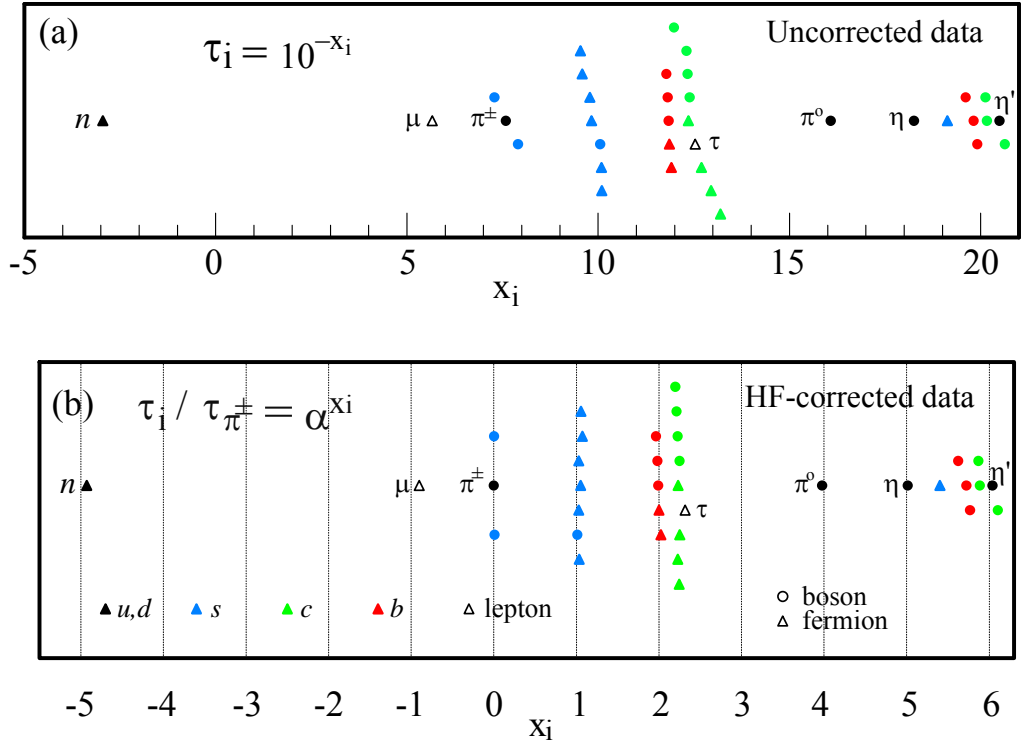

Fig. 1.  Logarithmic plots of the experimental lifetimes for the 36 long-lived particles with lifetime values $\tau > 10^{-21}$ sec (1 zeptosecond). *Fig. 1(a)* shows the lifetimes plotted as exponents $-x_i$ to the base 10 along the *x* axis, and spread out for clarity along the *y* axis. These lifetimes fall into well-separated groups that are determined by the dominant Standard Model quarks. The uniform slopes observed in the very-long-lived ($\tau > 10^{-14}$ sec) *s*-quark and *c*-quark lifetime groups are due to a factor-of-two lifetime "hyperfine" (HF) structure that is superimposed on the overall group spacings. *Fig. 1(b)* differs from Fig. 1(a) in three respects: *(1)* phenomenological "corrections" are applied to remove the HF structure; *(2)* the lifetimes are expressed as ratios to the $\pi^{\pm}$ reference lifetime; *(3)* these ratios are plotted as exponents $x_i$ to the base $\alpha = e^2/\hbar c$. As can be seen, the lifetime groups exhibit an accurate scaling in powers of $\alpha \cong 1/137$ that extends over 11 powers of $\alpha$. The lifetimes in the range $x_i \sim 0-2$ correspond to *unpaired* quark decays, and the lifetimes in the range $x_i \sim 4-6$ correspond to *paired* quark-antiquark decays or radiative decays. The paired-quark lifetimes are characteristically a factor of $\alpha^4$ shorter than the corresponding unpaired-quark lifetimes. The unpaired *c*-quark lifetimes are a factor of three shorter than the unpaired *b*-quark lifetimes. These $\alpha$-dependent lifetimes logically correspond to reciprocal $\alpha^{-1}$-dependent masses.

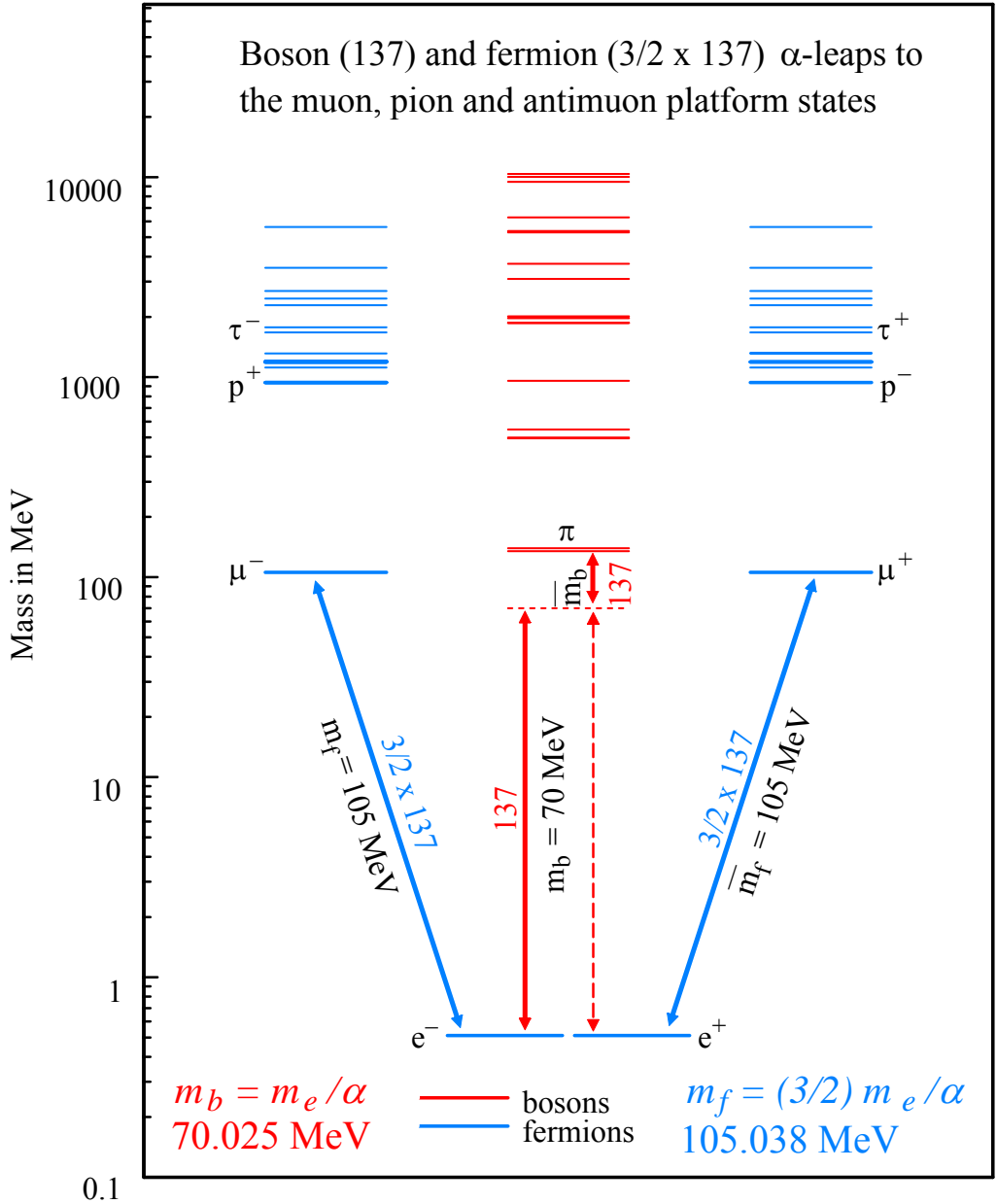

Fig. 2. A mass plot of the 36 long-lived particles of Fig. 1, together with the stable electron and proton. This plot shows two dominant mass ratios that contain 137 as a factor: *(1)* the $(m_b + \bar{m}_b) \cong 2 \times 70$ MeV α-leap from the electron + positron ground state to the 140 MeV $\pi \cong m_b \bar{m}_b$ boson; *(2)* the $m_f \cong 105$ MeV α-leap from the electron to the $\mu \cong m_f$ fermion. Relativistically, the mass units $m_b \cong 70$ MeV and $m_f \cong 105$ MeV occur as spin 0 and spin 1/2 configurations, respectively, of the same basic 70 MeV mass quantum.

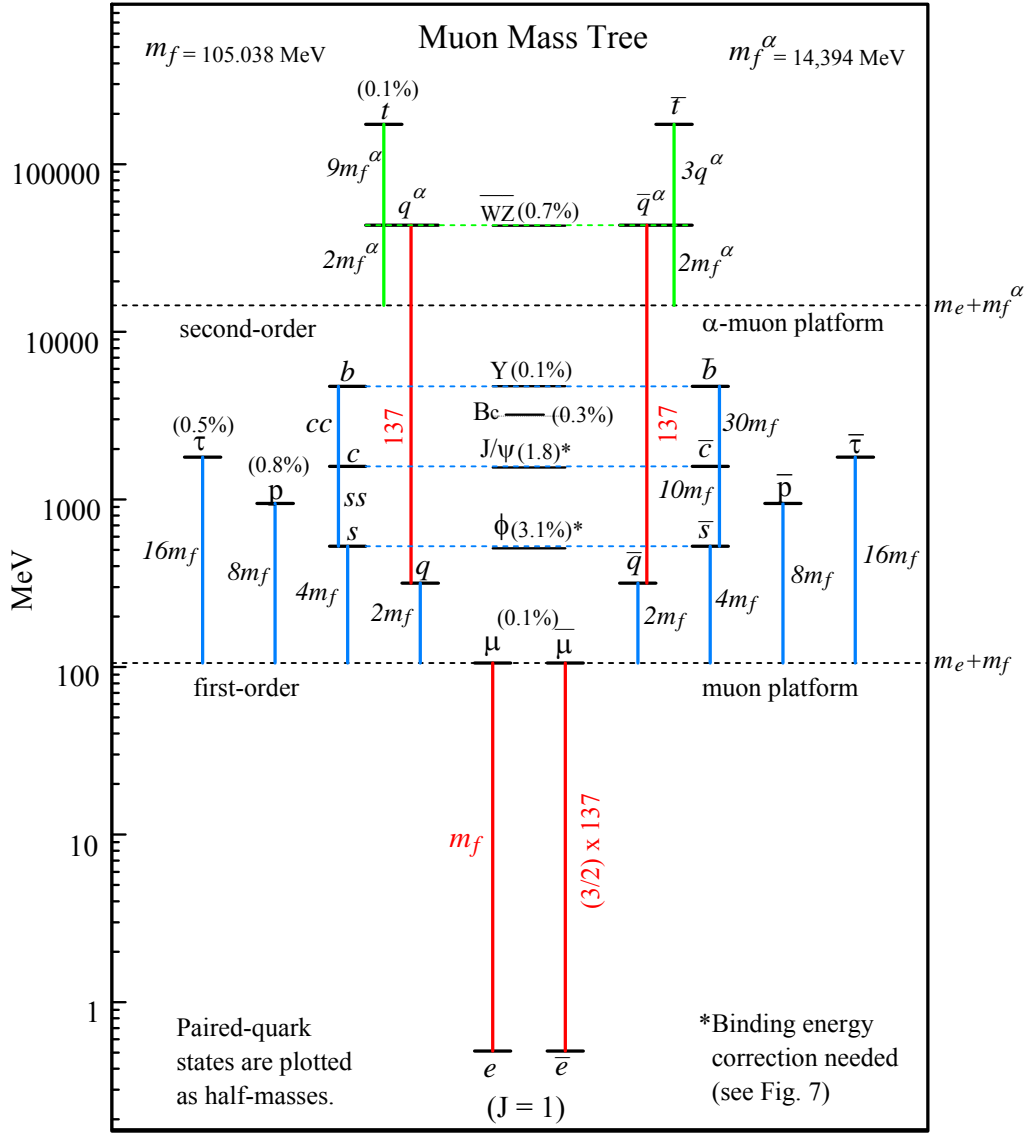

Figure 3. The spin 1/2 muon mass tree. An $\alpha$-leap of $m_f \cong 105$ MeV from an electron generates the fermion mass $m_f$, which combines with the electron mass to form the $m_\mu = m_e + m_f$ muon (0.1% accuracy). Then excitations of (2, 4, 8, 16) $m_f$ add to the muon platform mass to generate: *(1)* the $q \equiv (u,d)$ mass-averaged quarks; *(2)* the $s$ quark; *(3)* the proton p (with a double charge transfer); *(4)* the tau lepton $\tau$. Successive mass-triplings of the $s$ quark generate the $c$ and $b$ quarks (with a charge transfer at each step). The quark-antiquark pairs $s\bar{s}$, $c\bar{c}$, $b\bar{c}$ and $b\bar{b}$ reproduce the $\phi$, J/$\psi$, $B_c$ and $\Upsilon$ mesons. A second $\alpha$-leap of the $q \equiv (u,d)$ proton quarks by direct impact at the Tevatron generates the $q^\alpha \equiv q/\alpha$ quarks, which reproduce the mass-averaged $\overline{WZ}$ pair and the top quark $t$. The average accuracy of this additive constituent-quark mass systematics is at the 1% level (Table 1).

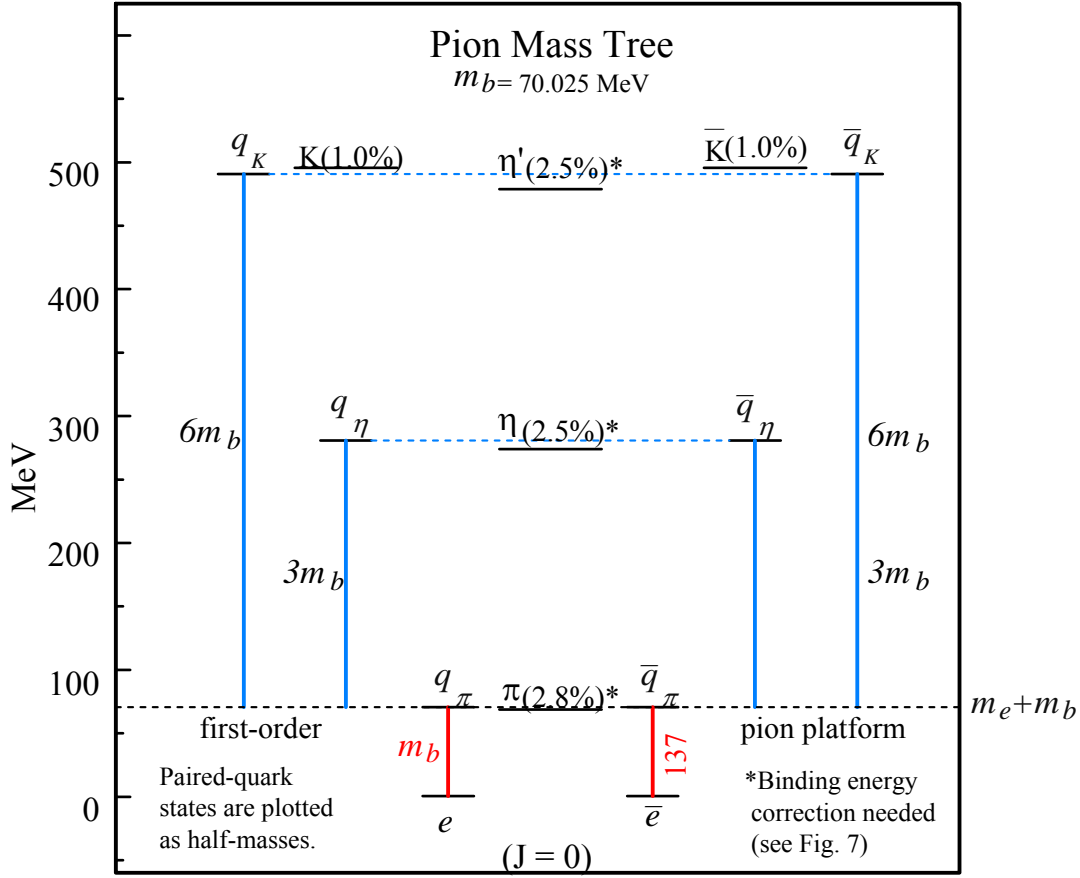

Figure 4. The spin 0 pion mass tree. An α-leap of $m_b \cong 70$ MeV from an electron generates an $m_b$ boson mass excitation unit. This combines with the electron mass to form a "generic" pion quark $q_\pi$ with the platform mass $m_{q_\pi} = m_e + m_b$. Then excitations of 3 $m_b$ and 6 $m_b$ are successively added to this $m_{q_\pi}$ platform mass to generate the generic $q_\eta$ and $q_K$ quarks. The quark-antiquark pairs $q_\pi \bar{q}_\pi$, $q_\eta \bar{q}_\eta$ and $q_K \bar{q}_K$ reproduce the π, η and η' pseudoscalar mesons, respectively. The masses of these three mesons each reflect a hadronic binding energy (HBE) of about 2.6%. The unbound $q_K$ mass matches the average kaon mass to an accuracy of about 1%. This pion mass tree reproduces the mass values of the pseudoscalar meson nonet of particles, but does not account for their isotopic spin values. The $q_\pi$, $q_\eta$ and $q_K$ quarks defined here are generic in the sense that we deal essentially just with their mass values, and do not specify their isotopic spin states.

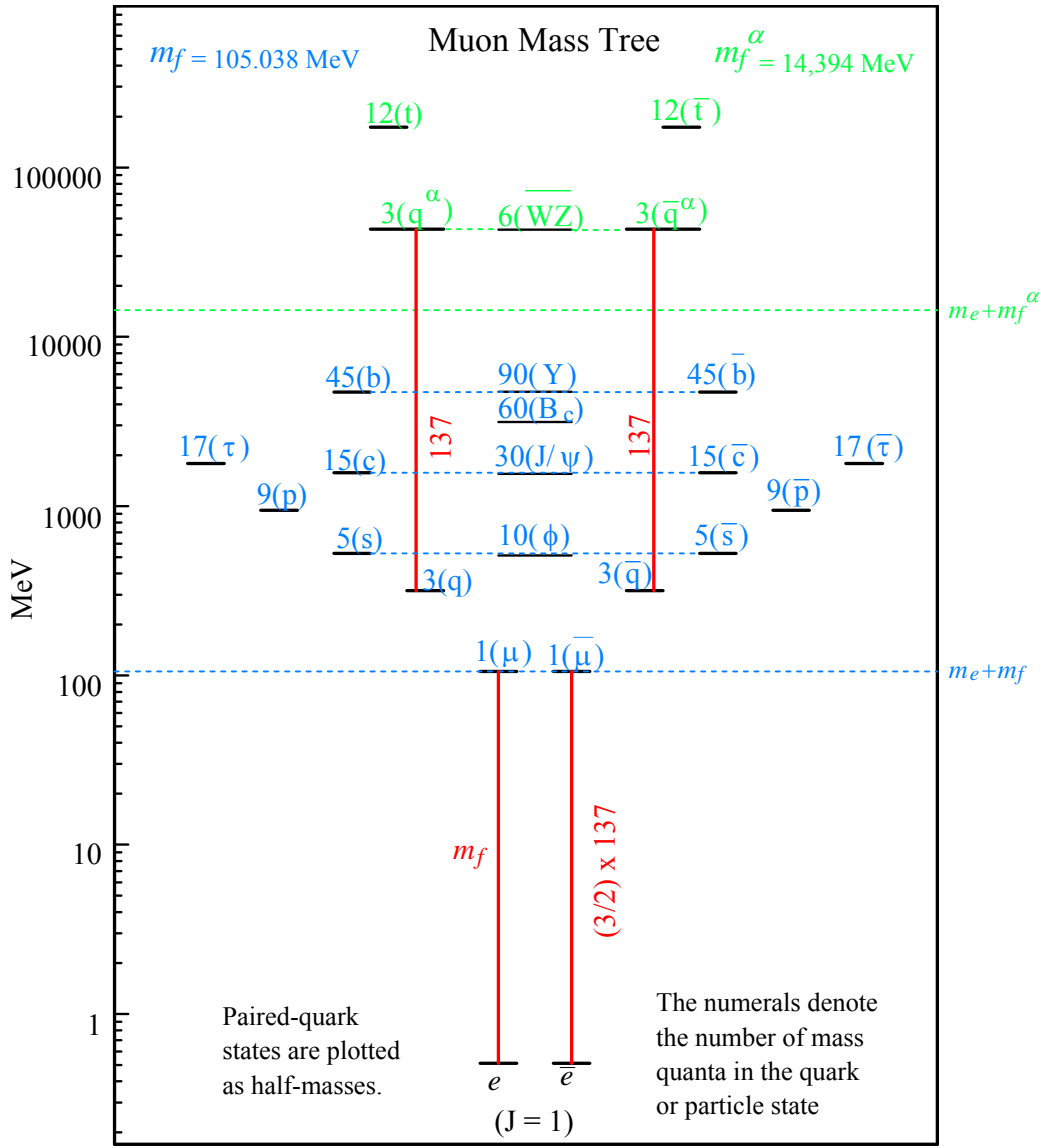

Figure 5. The muon mass tree of Fig. 3, shown plotted here with the number of $m_f$ or $m_f^\alpha \equiv m_f/\alpha$ mass quanta contained in each particle state. The masses extend from the lowest unstable particle, the muon, to the highest measured particle state, the top quark, and are all accurate to better than 1% except the hadronically bound and relatively-low-mass $\phi$ and $J/\psi$ vector mesons (Fig. 7 and Table 1). This plot constitutes a reply to Rabi's question about the muon: "Who ordered that?" Without the fundamental $m_f \cong 105$ MeV muonic mass quantum, this comprehensive mass framework would not exist. This muon-based framework, the "muon mass tree", interweaves leptons, hadrons, quarks and gauge bosons into a uniform mass pattern that requires all of these elements for its completion.

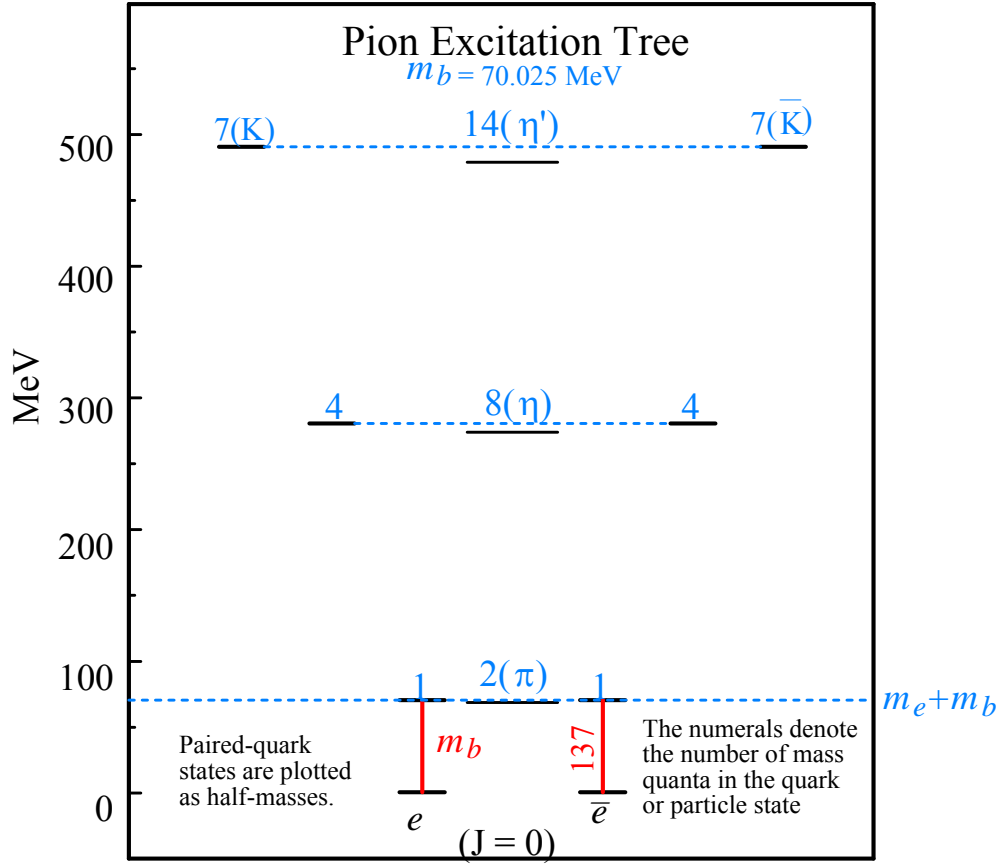

Figure 6. The pion mass tree of Fig. 4, shown plotted here with the number of 70 MeV mass quanta $m_b$ contained in each particle state. The particle states included here are the isotopic-spin-averaged members of the pseudoscalar (PS) meson nonet. The masses of the non-strange π, η, η' mesons follow a highly-accurate equal-interval rule, which indicates that the η and η' appear in this nonet on an equal footing. The PS lifetimes displayed in Fig. 1 also lead to this same conclusion. When paired-quark HBE mass corrections are applied (Fig. 7), the mass values for these states are reproduced at the 1% level (Table 1). The spin 0 mass quantum $m_b$ used here to create the pion mass tree does not exist in the Standard Model formulation. This pion mass tree is required in order to complement the constituent-quark systematics of the muon mass tree, which does not accommodate the PS meson nonet.

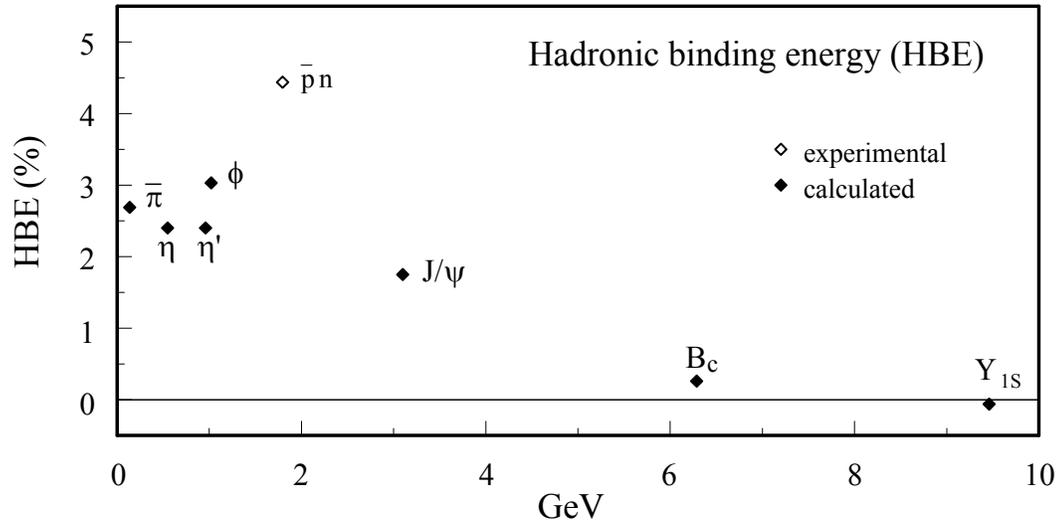

Fig. 7. Calculated hadronic binding energies (HBE's) for quark-antiquark particle states, shown together with the experimental binding energy of an antiproton-neutron pair as determined from its annihilation decay-product energy [9]. The calculated HBE's are obtained by using the quark masses defined in Table 1. As can be seen, the HBE is in the 2-3% range at low mass values and drops off to essentially zero at energies above 6 GeV. This drop-off at high energies can be qualitatively attributed to the decreasing Compton radii of the quarks with increasing mass, which brings the gluon fractional charge centers closer together and thus decreases the gluon "rubber band" forces, which is in line with the quark concept of "asymptotic freedom".